\documentclass[10pt,twocolumn,letterpaper]{article}

\usepackage{cvpr}
\usepackage{times}
\usepackage{epsfig}
\usepackage{graphicx}
\usepackage{amsmath}
\usepackage{amssymb}

\usepackage[breaklinks=true,bookmarks=false]{hyperref}

\cvprfinalcopy 

\def\cvprPaperID{****} 

\begin{document}

\title{Stretchcam: Zooming Using Thin, Elastic Optics}

\author{Daniel C. Sims\\
Department of Computer Science\\
Columbia University\\
New York, NY 10027, USA\\
{\tt\small dcs2140@columbia.edu}
\and
Oliver Cossairt\\
Department of Electrical Engineering\\
and Computer Science\\
Northwestern University\\
Evanston, IL 60208, USA\\
{}
\and
Yonghao Yue\\
Department of Complexity\\
Science and Engineering\\
Graduate School of Frontier Sciences\\
The University of Tokyo\\
Kashiwa-shi, Chiba, 277-8561, Japan\\
{}
\and
Shree K. Nayar\\
Department of Computer Science\\
Columbia University\\
New York, NY 10027, USA\\
{}
}

\maketitle

\begin{abstract}
Stretchcam is a thin camera with a lens capable of zooming with small actuations. In our design, an elastic lens array is placed on top of a sparse, rigid array of pixels. This lens array is then stretched using a small mechanical motion in order to change the field of view of the system. We present in this paper the characterization of such a system and simulations which demonstrate the capabilities of stretchcam. We follow this with the presentation of images captured from a prototype device of the proposed design. Our prototype system is able to achieve 1.5 times zoom when the scene is only 300 mm away with only a 3\% change of the lens array's original length.  
\end{abstract}


\section{Introduction}

\hspace{0.3cm} We are close to a world where surveillance can be ubiquitous. Multiple works have explored how to make cameras fully self-powered, which allow for cameras to be deployed without regard to access to a power source \cite{Nayar2015, Ahmed2016, Ko2017}. There has also been significant progress in low-power wireless streaming of data, with one example transmitting a 112 by 112 video at 13 frames per second 45.72 meters using only 2.36 mW \cite{Naderiparizi2017}. A recent paper described a system that was able to send data 2.8 kM using 9.25 $\mathrm{\mu W}$ \cite{Talla2017}. Ubiquitous imaging is near, but the power these easily deployable systems can produce is not abundant. These types of systems would need a new method of zooming to not add prohibitive power requirements to the design. traditional zoom lens requires large motors to move their relatively heavy components over long lengths of travel. As a result, they consume large amounts of power. Thus, there is a need for a simple mechanism for zooming that does not require a significant amount of motion.   


 
This paper takes a different approach to creating a zoom-capable imaging system. Our approach is to use a thin lens array that can change its field of view with barely perceivable motion in the form of a stretch. This reduction in travel distance offers not only an increase in zoom speed but power savings. 

 

\vspace{0in}
 
\begin{figure}[t]
 
\begin{center}
\includegraphics[width=0.7\linewidth]{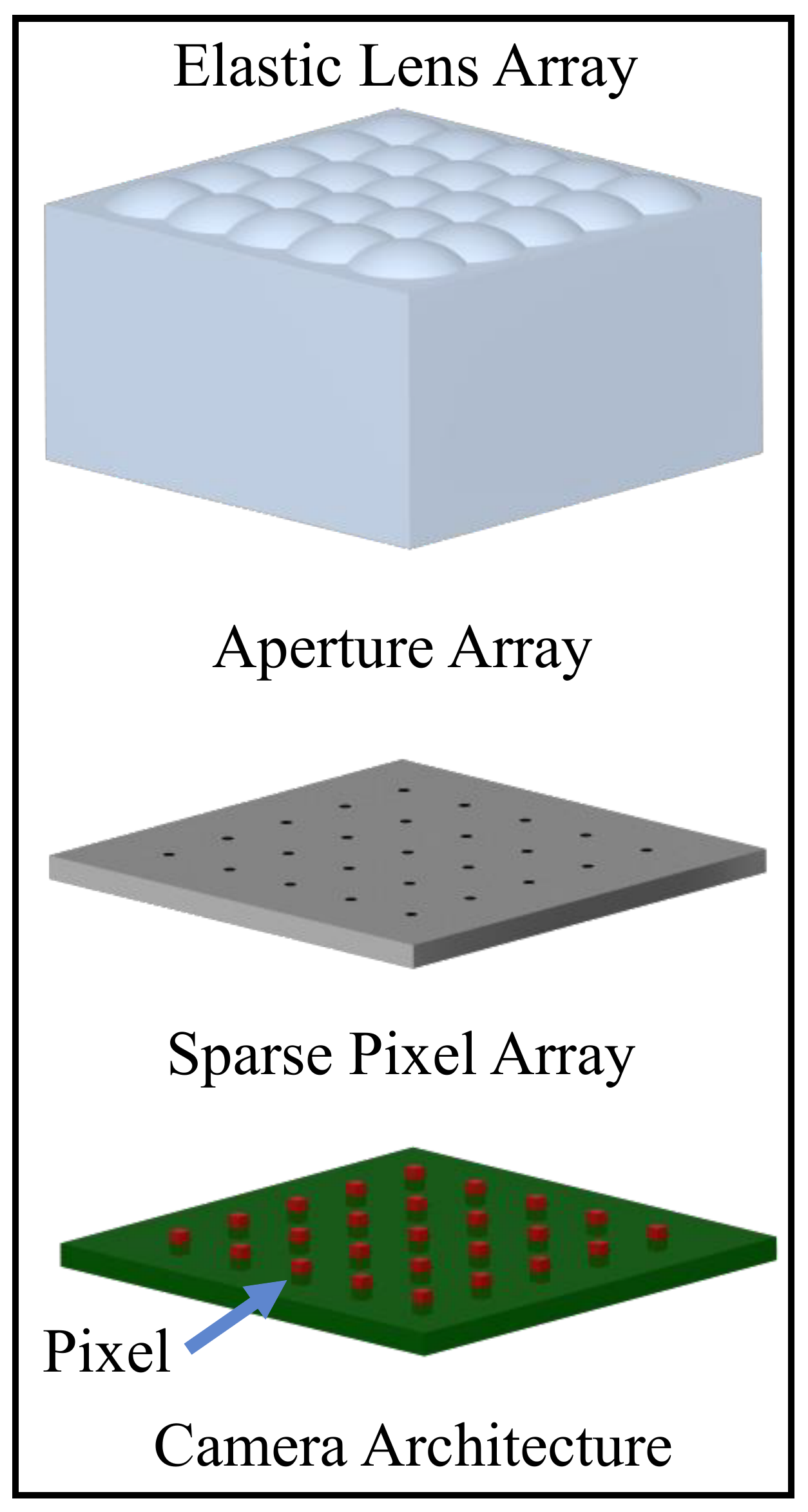}
\end{center}
\vspace{-0.00in}
\caption{Overview of the stretchcam. (a) A visualization of the camera architecture. An aperture array is placed on top of a sparse pixel array, then an elastic lens array, whose pitch is equal to the pitch of the sparse pixel array is placed on top of the aperture plate. }
\vspace{-0.15in}
\label{fig:overview}
\end{figure}

Stretchcam's architecture is an elastic lens array, such as in Fig. \ref{fig:overview}, placed on top of an aperture sheet that is covering a rigid, sparse pixel array. The pitch of the pixel array is the same as the lens array. Each lens in the array focuses light onto its corresponding pixel. 



%

\begin{figure*}[t]
\begin{center}
\includegraphics[width=0.75\linewidth]{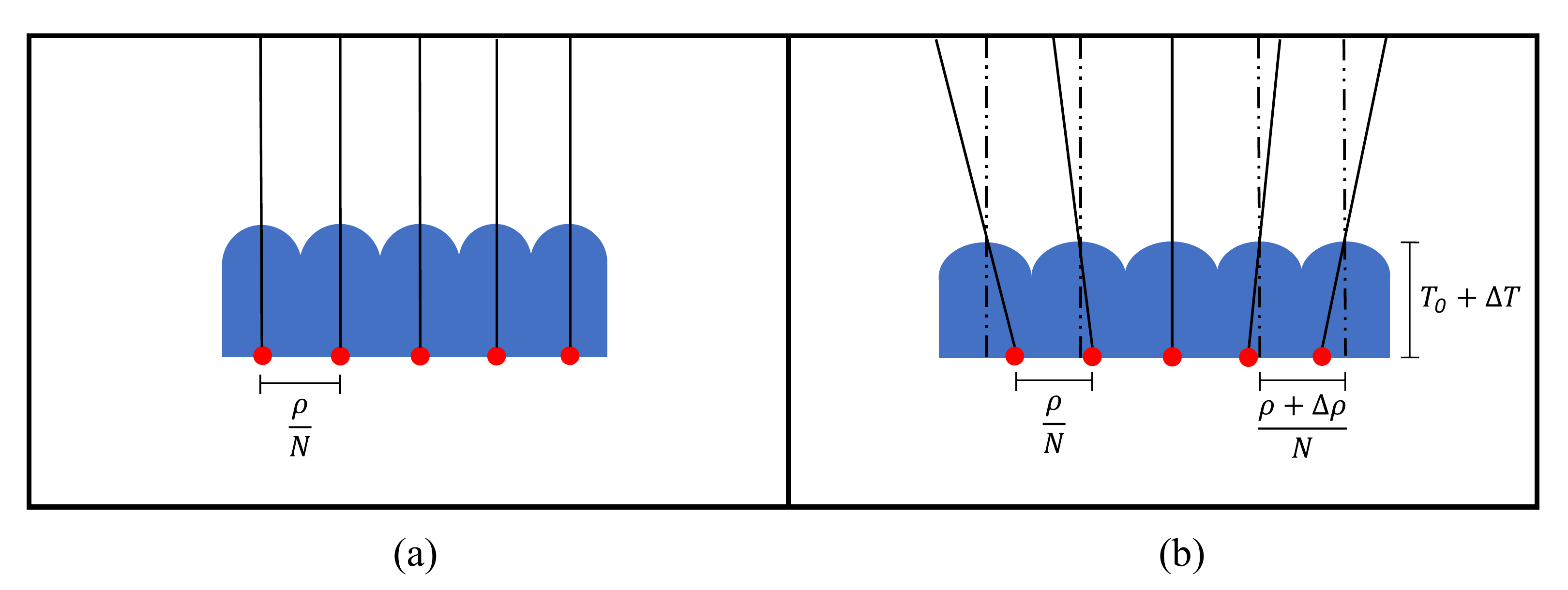}
\end{center}
\vspace{0.0in}
\caption{The unstretched and stretched configurations of Stretchcam are presented in this figure. The lens array (blue) placed on top of a sparse array of pixels (red) whose pitch is equal to the lens array's. Stretching the lens will change the angle between the primary ray and the optical axis of each lens (black line). In (a), the pixels are coincident with the optical axis of its corresponding lens and, in turn, the primary ray for each lens in the array is coincident with the optical axis of its corresponding lens. In (b), the elastic lens array is stretched a small amount causing an increase in field of view of the imaging system. }
\vspace{-0.0in}
\label{fig:base_idea}
\end{figure*}

To understand how stretching an elastic lens would change the field of view, consider the scenarios illustrated in Fig. \ref{fig:base_idea}. In case \ref{fig:base_idea}(a), the pitch of the sparse pixel array and the lens array is $\frac{\rho}{N}$, where $\rho$ is the undeformed lens array width and $N$ is the number of lenses in the array. In case \ref{fig:base_idea}(b) the lens array is stretched a small amount, $\Delta\rho$; the pitch of the sparse pixel array remains $\frac{\rho}{N}$ but the pitch of the elastic lens array increases to $\frac{\rho + \Delta \rho}{N}$. This increase in the lens array's pitch while the pitch of the pixel array remains constant tilts the primary ray of each lens in the array. Elastomers conserve their volume when deformed, thus the thickness of the lens array, $T_0 + \Delta T$, decreases as the lens array is stretched, further increasing the tilt of the primary rays. The combined effect is that Stretchcam's field of view increases a substantial amount with a small deformation.
 
The primary contribution of this paper is a camera that uses small deformations (approximately 3\% of the total lens array length) of an elastic lens array to achieve large zoom (1.5 to 2 times zoom), the characterization of such a system and the presentation of images captured from a prototype device of the proposed design. We also contribute the mechanical principles and the optical properties of Stretchcam as well as several simulations to verify the efficacy of the proposed design. 

The key novelty of Stretchcam is that it allows for the creation of thin, zoom capable imagining systems using cheap materials and simple mechanics. Thus, the outlined approach has the potential to undercut the costs and power demands of other proposed methods while achieving fast zoom times. 

\vspace{-0.1in}
\section{Related Work}

 
Several other authors have explored various methods to overcome the large travel distances and power demands of some zoom lenses on the market. For example, electrowetting \cite{Li:16,Peng:07} -- a phenomenon where the electric charge on a plate will impact the amount a droplet of liquid spreads on a surface -- is one method that has been rigorously explored to create zoom lenses. This approach allows for electronic control of a lens' shape; however, as Graham-Rowe \cite{Graham-Rowe2006} points out, gravity will impact the performance of these lenses. There is also a limited number of liquids suited to the purpose, and this limit restricts the optical properties available to designers. 

Lin et al. \cite{Lin11} built an electrically tunable zoom system with two liquid crystal lenses, however, they report a four-second response time. Blum et al. \cite{Blum2011} acknowledge the slow response times of liquid crystal lenses and instead use mechanically actuated polymer lenses to create a zoom lens. Deformable optics \cite{Carpi2011,Beadie:08} can make a single aperture zoom system thin by eliminating the need for many rigid optical components. Stretchcam, however, uses the properties of lens arrays to create a thin imaging system and adds a zooming capability.

Folded zoom lenses, which use prisms or mirrors to angle the optical axis to make the camera thinner, have been proposed \cite{Reiley2014,Yabe2015}. Light \cite{Light} is a commercially available camera that uses multiple folded optics to create a form factor that is only 24 mm thick yet can capture a 52-megapixel image. It features a five times optical zoom achieved by tilting the mirrors in the system. Folded zoom lenses are interesting because they retain the properties conventional lenses while making the system thin and thus start to address the bulk issue mentioned in the introduction, but leave the expense involved in such systems, slow zoom times and power demands unaddressed. Lu et al.\cite{Lu2012} extend the idea of using mirrors in zoom lens by using rigid and MEMS deformable mirrors to create a thin, zoom capable optical system. The team is able to eliminate the complex moving components along with glass optics with their electronically deformable mirror. 

Multiple aperture systems, like lens arrays, can outperform single aperture systems in terms of thickness. One can reduce the thickness of a single aperture system by decreasing the focal length of the lens. To maintain the same $f$-number there must be a corresponding decrease in the diameter of the lens. However, as Lohnman \cite{Lohmann89} explains, the space-bandwidth product decreases as the scaling factor of the lens decreases; very few pixels are resolved in a thin, single aperture system. 

V{\"o}lkel et al. \cite{Volkel2003} discuss how using a multiple aperture system in the form of a lens array can overcome this barrier. Lens arrays were used by Portnoy et al. \cite{Portnoy09} to reduce the thickness of an infrared camera. Br{\"u}ckner et al. \cite{Bruckner2011} designed a 1.4 mm thick camera that can achieve VGA resolution and a 2 mm thick camera that achieved 720p resolution \cite{Bruckner2014} with both designs using a lens array to achieve their thinness. Oberd{\"o}tster et al. \cite{Oberdorster12} had a similar design that could change its front focal point by compensating for different amounts of parallax. Venkataraman et al. \cite{Venkataraman2013} also use a lens array to make a 3.5 mm thick camera that has similar image quality to the iPhone 5. These, however, are all rigid lens arrays without optical zooming.
 
Tanida et al. \cite{Tanida01} proposed a design where a lens array is used to achieve a thin optical configuration. They can manipulate the position of their photo detectors to capture a demagnified image of the system and use a back-projection method to generate an image that is a higher resolution than the lens array used to capture the scene. Sims et al. \cite{Sims16} bend a lens array to vary its field of view and design it such that it avoids aliasing.
 
This paper looks at stretching as a mechanism for zooming. Stretching an elastic lens array to change its optical properties has been used to improve various aspects of integral imagers. Kim et al. \cite{Kim08} proposed a design where a lens array is fitted over a sensor to capture an integral photograph of a scene. The lens array is then stretched to match the pitch of a display with a different pitch than the sensor. This ability to match the pitch of the lens array eliminates artifacts in the integrated image. Kim et al. \cite{Kim16} use a stretchable lens array as a component in a light field microscope to match the image side $f$-number of the main objective lens to use the full resolution of an image sensor. 

Other researchers have built systems with deformable lens arrays to tune the focal length. Li et al. \cite{Li2015118} conducted a finite element analysis of a lens array placed in tension and studied the corresponding increase in focal length of the lenses in the lens array. Chandra et al. \cite{Chandra07} built concave and convex elastic microlens arrays whose focal lengths are tunable by stretching. Stretchcam, however, is designed to achieve the same zooming as a conventional lens but with small mechanical actuations and in a thin form factor.  
 
\section{The Mechanical Principles of Stretch Imaging}
 
 
 
\hspace{0.3cm} Stretchcam consists of a rigid plate with a sparse array of pixels. On top of this rigid plate, there is an elastic lens array with the same pitch as the pixel array. In order to deform the lens array, a mechanism is placed on either side of the camera and attached only to the elastic lens array. No other deformations are allowed in Stretchcam except for the stretching of the lens array. However, the lens array is attached to the mechanism such that its thickness will be unconstrained. As explored in the literature \cite{Kim08,Kim16,Li2015118,Chandra07}, when a lens array is stretched the shape of the individual lenses in the array will change. This change in shape will impact the imaging properties of the lenses. This stretching will also impact the viewing direction of each lens in the array, which in turn demagnifies the scene.

\begin{figure*}[t]
\begin{center}
\includegraphics[width=0.75\linewidth]{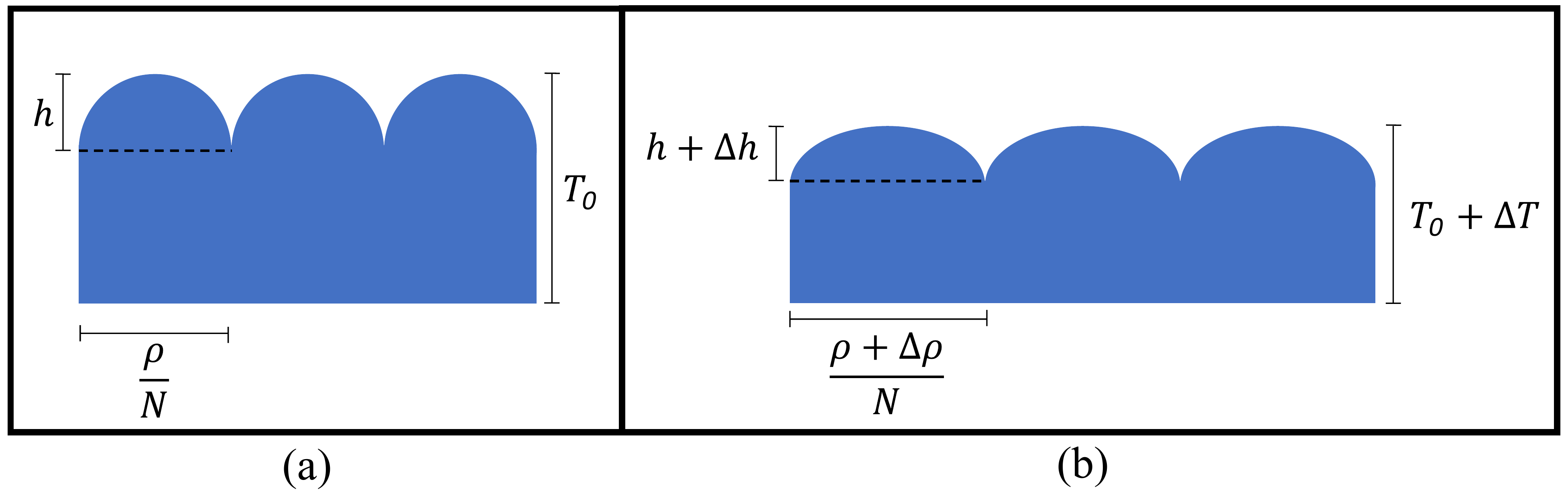}
\end{center}
\vspace{-0.2in}
\caption{A cross-sectional view of a lens array undergoing deformation. Notice how the spherical portion of the lens array flattens in (b) as the amount the lens array is stretched by increases. }
\vspace{-0.20in}
\label{fig:Lens_shape}
\end{figure*}

Since the lens array is made of a silicone elastomer it will conserve its volume when deformed which means that stretching the lens array changes its thickness. $T(\Delta \rho)$, the change in the lens array's thickness as a function $\Delta \rho$, the amount the lens array is stretched, is defined as

\begin{align}
\label{eq:stretch_vs_thickness_function}
T(\Delta \rho) = \frac{\rho}{(\rho + \Delta \rho)}T_0
\end{align}
 
\noindent where $T_0$ is the undeformed lens array thickness and $\rho$ is the undeformed lens array width as depicted in Fig. \ref{fig:Lens_shape}.
 
The radii of the individual lenses is a function of the lens array thickness and the width of the individual lenses in the array. As long as the lens array is stretched a small enough amount such that the lenses remain spherical, then the chord length and sagitta of the circular segment that comprises each lens in the array can be used to find the radius for the individual lenses in the array. This radius as a function of the amount the lens is stretched, $R(\Delta \rho)$, is given by
\begin{align}
\label{eq:deformed_radius_rho}
R(\Delta \rho) = \frac{h\rho^2}{2(\rho+\Delta \rho)^2}+\frac{(\rho + \Delta \rho)^4}{8N^2h^2\rho^2}
\end{align}
 
\noindent where $N$ is the resolution of the lens array and $h$ is the initial sagitta length that makes up the circular segments of the lenses as indicated in Fig. \ref{fig:Lens_shape}.
 
Since the sensor is attached directly to the bottom of the plano-convex lens, the back focal distance is the lens array's thickness. The refraction that occurs between the bottom surface of the lens array and the pixel is ignored. When this assumption is true, only the focal length of the individual spherical surfaces in the lens array need be determined to model the system. The front focal length as a function of the amount the lens is stretch, $d_f(\Delta \rho)$, is therefore
 
\begin{align}
\label{eq:refraction_at_spherical_surfaces}
d_f(\Delta \rho) = \frac{1}{\frac{\eta-1}{R(\Delta \rho)}-\frac{\eta}{T(\Delta \rho)}}
\end{align}
 
\noindent where $\eta$ is the index of refraction for the elastic material. The direction each lens is viewing the scene is also a critical aspect of the system to understand. The angle between any lens' primary ray and optical axis, $\theta_i$,  is given by
 
\begin{align}
\label{eq:shift}
\theta_i = arctan(\frac{\Delta \rho i}{N(T_0 - \Delta T)})
\end{align}
 
where $i$ is the index of any lens in the array. The center lens has an index of zero if there is an odd number of lenses in the array whereas the two most extreme lenses in the array have an index of $\pm \frac{N-1}{2}$. The field of view of Stretchcam as described by the width of the scene imaged by the camera in one dimension, $s$, is then

\begin{align}
\label{eq:stretch_cam_field_of_view}
s = \frac{(N-1)}{N}(\rho+ \Delta \rho + \frac{\Delta \rho d}{T_0 - \Delta T}).
\end{align}

Equation (\ref{eq:refraction_at_spherical_surfaces}) describes how individual lens react to the deformation and Eq. (\ref{eq:shift}) describes the direction any individual lens in the array views the scene as the lens is deformed. With these two equations, it is possible to determine if a desired scene is captured without unacceptable blurring or aliasing not only for any static case but throughout the deformation.
 
\section{The Optical Properties of Stretch Imaging}
 
\subsection{Focal Plane of a Lens Array}

\begin{figure*}[t]
\begin{center}
\includegraphics[width=0.40\linewidth]{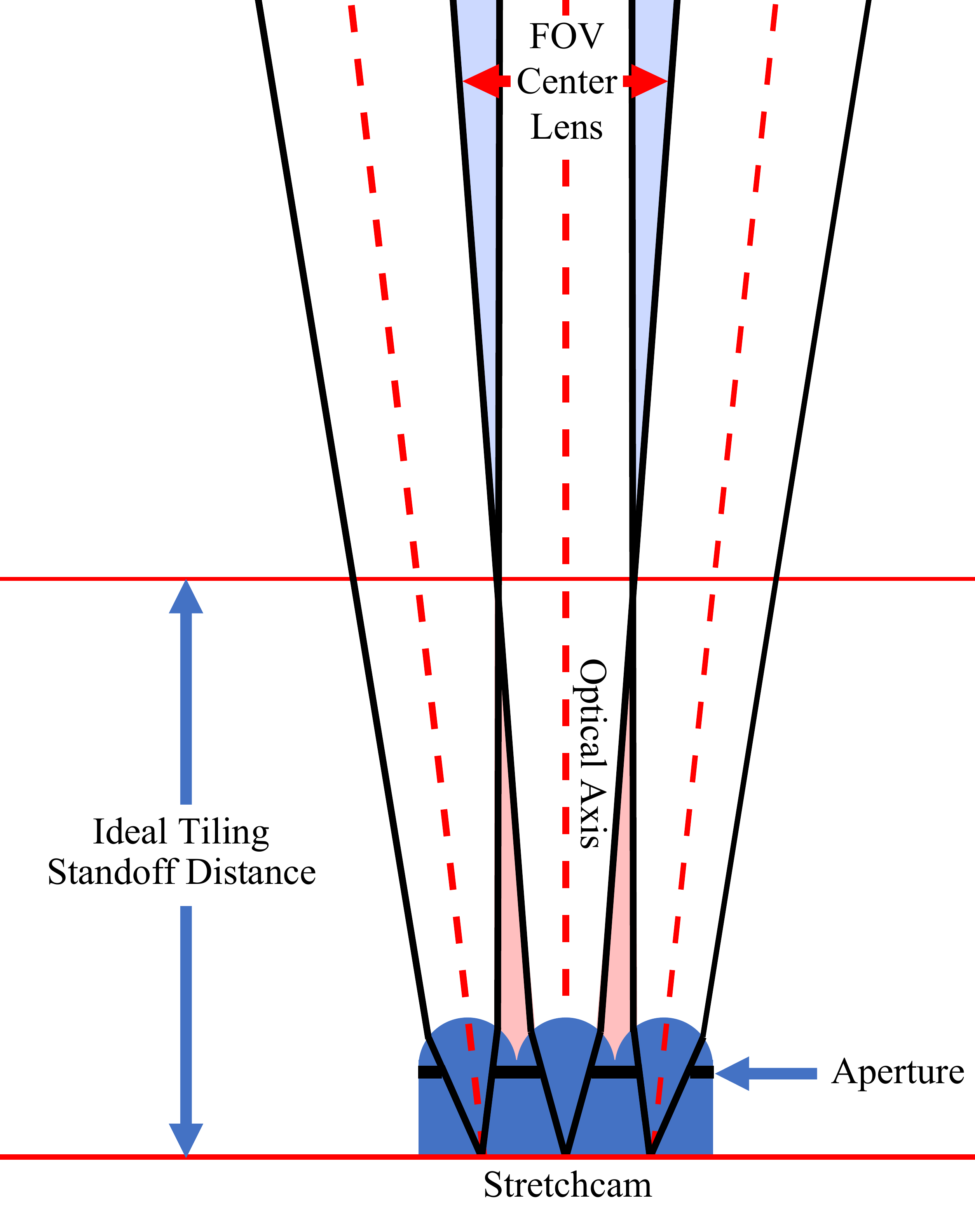}
\end{center}
\vspace{-0.15in}
\caption{Ideal Tiling. For an ideal lens overlaps in the captured patches - which cause blurring - occur in the near and far field. For Stretchcam,  gaps - which cause aliasing - occur in the near-field as indicated by the red transparent regions and blurring occurs in the far-field as indicated by the blue transparent regions. There is a standoff distance for which the system is in focus. This is when the patches each pixel captures are ideally tilted. }
\vspace{-0.25in}
\label{fig:Ideal_tiling_blurring_and_Alaising}
\end{figure*}
 
 
\hspace{0.3cm} For an ideal lens, each pixel on a sensor will integrate the light in the area of the scene. This area is determined by the projection of that pixel through the center of projection onto the scene. The system is in focus when these regions abut which happens at the ideal tiling standoff distance. If the scene is not at the focal plane the patches each sensor captures would get wider and would no longer abut, this would cause the imaged regions to overlap and in turn case the final captured image to be blurred. 


 
Stretchcam's focal plane is defined in the same way as a conventional camera's: the plane for which the patches each pixel captures is ideally tiled, with no overlaps or gaps between the regions. Scenes captured by the system will be blurred if in the far-field, as indicated by the blue shaded regions in Fig. \ref{fig:Ideal_tiling_blurring_and_Alaising}, but the captured images will be aliased if the scene is near the lens and the apertures of the individual lenses are not equal to the width of the individual lenses. This is indicated by the red shaded regions in the same figure. If the aperture is equal to the individual lens width, then captured images will not have aliasing artifacts in the near-field but instead have blurring in the near field.

For simplicity, this version of Stretchcam's design has the focal length of the individual lenses in the array will equaling or exceeding the thickness of the lens array. This assumption is not burdensome. The lens array can be designed to be in focus at the desired standoff distance in the undeformed case such that the focal length of the individual lenses is greater than the lens array thickness. If the lens array is designed this way, the assumption is true throughout the stretch. With this set of assumptions, region of the scene the individual lenses in the array capture at arbitrary distance between the scene and the camera, $d$, is
 
\begin{equation}
\label{eq:patch_size}
a_w = \frac{(P_w|d_f|+\frac{\rho + \Delta \rho}{N}(T_0 + \Delta T))d}{2(T_0+\Delta T)|d_f|}
\end{equation}
 
\noindent where $P_w$ is the pixel width. With this it is possible to determine if the imaged patches do not abut by finding the sign of $a_w(\Delta \rho , d) - s(\Delta \rho , d)/N$. If this value is positive, there is aliasing due to gaps between the imaged regions; if this value is negative, there is blurring due to overlap in the patches; and if the value is zero, then the system is in focus. While both $a_w$ and $s$ increase as functions of standoff distance $d$, $a_w$ increases much faster than $s$. This means that the ideal tiling standoff distance will increase.

\subsection{Maximizing Spatial Resolution}
 
\subsubsection{Mechanical Limits on Resolution}
 
\hspace{0.3cm} The architecture of Stretchcam imposes several limits on resolution. One source of these limits is the ideal tiling condition. To achieve a clear image of the scene, the patches that the individual pixels of Stretchcam capture must abut. Thus, there is a limit to the number of captured patches that can fit within the camera's field of view. To fit more captured regions within the camera's field of view, the patches must be made smaller. To make these patches smaller, the initial lens array thickness, $T_0$, must increase or the pixel width, $P_w$, must decrease according to the pinhole model.
 
 
The size of an ideally tiled patch is $\frac{S}{N-1}$ where $S$ is the width of the scene captured by the lens array at desired standoff distance $D$. There is a limit to the number of patches that can fit on a scene, $N_{C1}$, that is defined by
 
\begin{equation}
\label{eq:resolution_constraint_one}
N_{C1} \leq \frac{S*T_0}{P_w*D} + 1
\end{equation}
 
\noindent where $D$ is the distance to the scene. To maximize this criterion, the pixel size and standoff distance must be as small as possible. The field of view and the initial lens array thickness must be as large as possible. This is a problem as, recalling Eq. (\ref{eq:stretch_cam_field_of_view}), the change in the field of view for the same amount of strength is greatest in the thinnest possible lens. Thus, the resolution may need to be sacrificed to gain greater change in resolution.

This is not the only limit on the resolution of the lens array. If the difference between the pixel array's pitch and the elastic lens array's pitch is too large, some pixels may be positioned under an adjacent lens resulting in crosstalk. Crosstalk will happen even in this case if the lens array is idealized as having thin baffles between each lens. This additional limit, $N_{C2}$, is defined as

\begin{equation}
\label{eq:resolution_constraint_two}
N_{C2} \leq \frac{\rho*(D-T_0)}{T_0*(S-\rho)}.
\end{equation}

Making the lens thinner increases the Stretchcam's resolution that meets this criterion. However, the first resolution constraint, Eq. (\ref{eq:resolution_constraint_one}), is increased by making the lens thicker. There exists a thickness that will give the highest resolution that is valid for both $N_{C1}$ and $N_{C2}$. Further, $N_{C2}$ is the highest resolution in the undeformed case that still avoids crosstalk. The resolution of Stretchcam must be lower than this number if the user is to stretch the system without causing crosstalk. We can calculate the approximate resolution criteria in the stretched case by assuming that that thickness changes a small amount. 

Take, for instance, a version of Stretchcam that is 80.475 mm wide. The lens array is 2 mm thick and each pixel is 0.001 mm wide. The width of the scene in the stretched case is 149 mm wide. The thickness of the lens does not change a large amount during the expected 0.555 mm stretch, so we can assume that $T_0 \approx T_0 + \Delta T$. In this case, $N_{C1} = 1288$ and $N_{C2} = 145$. Thus, the maximum resolution that Stretchcam can achieve in this configuration is 145 by 145 pixels. 

 
\subsubsection{Diffraction Blur Limits on Resolution}
 
\hspace{0.3cm} Because each lens in Stretchcam has an aperture whose size is less than or equal to the width of the individual lenses in the array, increasing the resolution increases the diffraction blur because the apertures must get smaller. There comes a point where attempting to make the lenses in the array smaller to increase the resolution would fail to improve image quality, assuming that the width of the lens array is constrained. This balancing point comes when the field of view of an individual lens in the undeformed case is equal to the angular separation needed to see two distinct object points due to diffraction or
 
\begin{equation}
\label{eq:resolution_constraint_three}
\frac{S}{ND} \geq \frac{1.22\lambda N}{\rho}
\end{equation}

\noindent where $\lambda$ is the wavelength of light captured by the system. This resolution limit, $N_{C3}$, can be defined as
 
\begin{equation}
\label{eq:resolution_constraint_three}
N_{C3} \leq \sqrt{\frac{S*\rho}{1.22*\lambda*D}}.
\end{equation}
 
Take, for instance, light at a wavelength of 600 nm illuminating a scene imaged by the previously mention version of the Stretchcam. For this system, the maximum resolution that still meets this constraint is 186 by 186 pixels. While this constraint is still above the $N_{C2}$ constraint that is limiting the example system, this constraint must be considered. Simply moving the desired focal plane to 500 mm drops $N_{C3}$ to 131 pixels. Whereas $N_{C1}$ becomes 322 pixels at one meter and $N_{C2}$ becomes 498 pixels at one meter.

\subsection{Effect of Lens Array Thickness on Change in Field of View}
 
\hspace{0.3cm} Instead of optimizing the resolution, some designers may wish to maximize the change in the field of view produced by stretching the lens. This can be done by either decreasing the thickness or increasing the width of the individual lenses. Decreasing the thickness makes it so stretching the lenses causes a larger change in the field of view. Increasing the width of the individual lenses allows the lens array to be stretched more before cross talk occurs. If the lens array's width is constrained, increasing the width of the individual lenses will reduce the resolution. Thus optimizing for the change in the field of view is best done by reducing the thickness of the lens array.
 
However, adjusting the thickness alone would cause the rays leaving the lens, which were previously parallel to the optical axis, to diverge. In order to correct this, a lens with a shorter focal length is used. An aperture is added in order to reduce spherical aberrations. This has the effect of adding aliasing in the near-field.

\subsection{Nyquist Frequency of Elastic Lens Arrays}
 
\hspace{0.3cm} In addition to the imaged regions of the scene not abutting, the Nyquist frequency of Stretchcam is a factor that will impact the quality of captured images. The quality of captured images will decrease as the lens is stretched. This is due to the increasing period between each sample on the scene as the lens is stretched. The Nyquist frequency of the Stretchcam as a function of $\Delta \rho$ is

\begin{equation}
\label{eq:Nyquist_Frequency}
\frac{1}{2}\frac{N* T(\Delta\rho)}{\Delta\rho * D + (\rho + \Delta\rho)*T(\Delta\rho)}.
\end{equation}

Consider the example system discussed in Section 4.2.1 when $\Delta\rho$ equals $0$, $0.2775$ and $0.555$. At a standoff distance of 250 mm the corresponding Nyquist frequencies are 0.9 lp/mm (Line Pairs per mm), 0.627 lp/mm and 0.481 lp/mm. As expected, the Nyquist frequency decreases as the lens is stretched. Thus, there will be a corresponding decrease in image quality. The Nyquist frequency remains 0.9 lp/mm when the standoff distance is 500 mm and the system is unstretched since every individual lens' primary ray is parallel with each other. However, when $\Delta\rho$ equals $0.2775$ and $0.555$ the Nyquist frequency is only 0.482 lp/mm and 0.328 lp/mm, respectively. This trend continues, resulting in increasingly worse Nyquist frequencies in the stretched cases as the standoff distance increases. 

 
\subsection{Impact of Free Variables}

We consider a version of Stretchcam that is focused on a scene that is 250 mm away. The lens array is approximately 80.5 mm by 80.5 mm and the desired demagnification amount is 1.85 times. It is desirable to increase the resolution of the system and reduce the amount the lens array needs to be stretched in order to change the magnification. Consider Table \ref{table:compare}, for a Stretchcam of any design, reducing the pixel width and making the lens array thinner will improve both the optimal resolution and the max value of $\Delta\rho$ needed to change the amount of magnification. This shows a key benefit of the design of stretchcam, making the camera thinner improves the performance of the camera.

 \begin{table*}
\begin{center}
\includegraphics[width=0.95\linewidth]{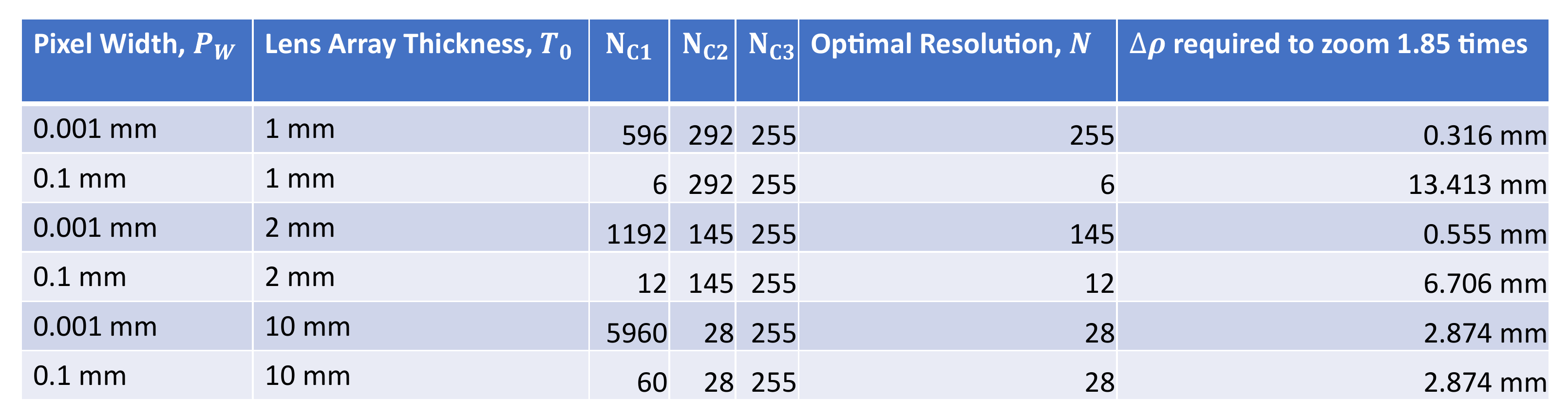}
\end{center}
\vspace{-0.15in}
\caption{Impact of Free Variables. This table shows the three resolution criteria, the optimal resolution and max $\Delta\rho$ for various values of lens array thickness, ${T_0}$, and pixel width, $P_W$. Making the lens array thinner and the pixel width smaller increases the optimal resolution and decreases the amount the lens array needs to be stretched in order to zoom.}
\vspace{-0.10in}
\label{table:compare}
\end{table*}

\section{Simulated Performance}
 
\hspace{0.3cm} Using the model of our Stretchcam, we conduct several simulations to evaluate its optical and mechanical properties further.
 
We first conduct a simulation using the finite element analysis package Abaqus \cite{Abaqus} to determine if the lens array will stretch evenly. In this simulation, we model a 7 by 7 lens array that is 49 mm on either side and 23 mm thick. Each lens has a radius of 7.5 mm. We simulate stretching the lens array in one dimension by 7 mm in a way that emulates gluing the two edges of the lens array to a vice-like mechanism, then spreading the jaws. We used C3D10 elements in the simulation and used the Neo Hooke model. We assumed the bulk modulus of our material to be 1.804 and the shear modulus to be 0.034. The horizontal boundaries are unconstrained, causing those boundaries to bow inwards.
 
\begin{figure}[t]
\begin{center}
\includegraphics[width=0.85\linewidth]{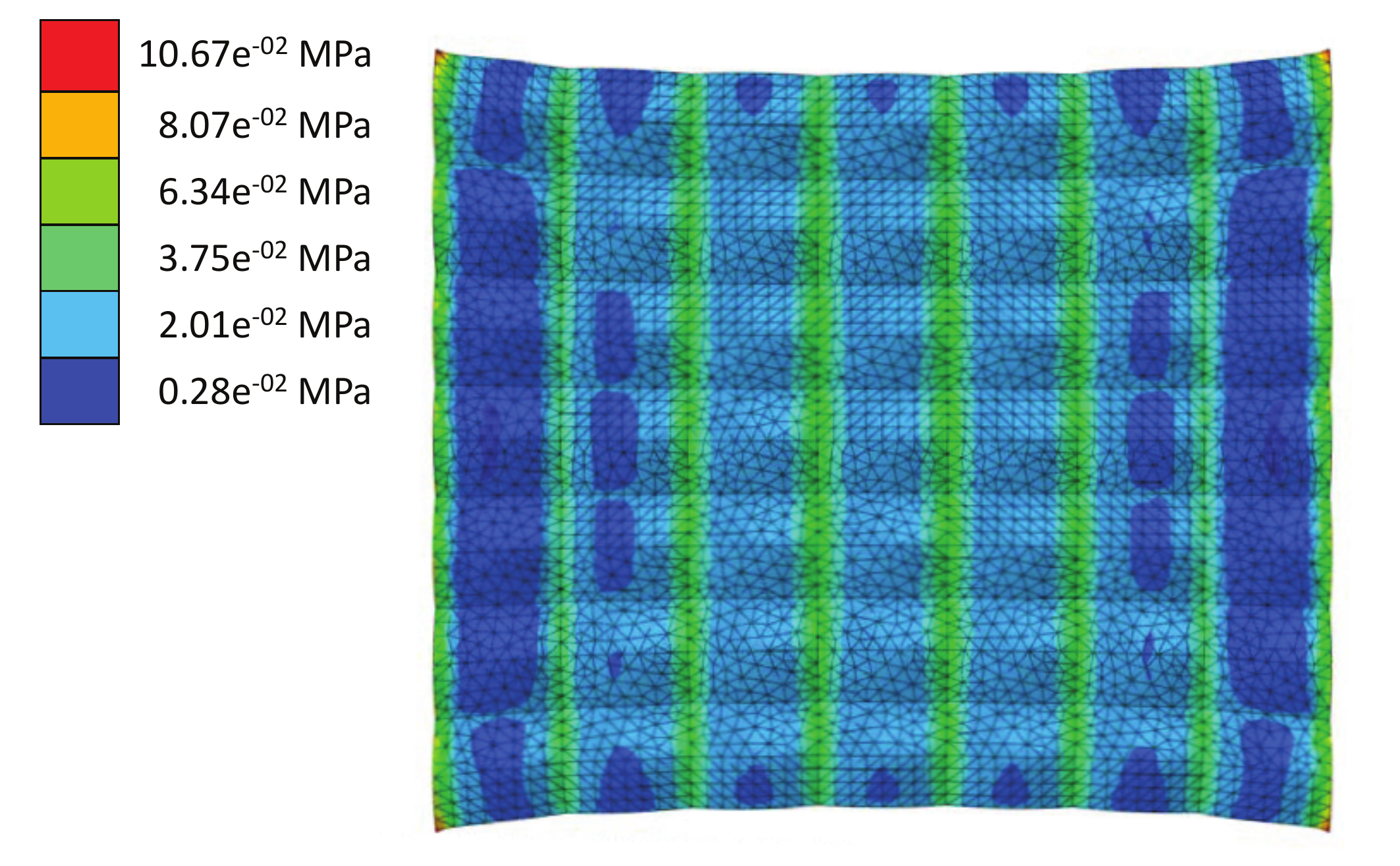}
\end{center}
\vspace{-0.15in}
\caption{Abaqus Simulation. This is a Abaqus simulation of a 7 by 7 lens array with 7 mm by 7 mm wide lenses stretched by 1 lens length. While there is unevenness in the stress distribution at the edge of the array, the 3 center columns and 5 center rows all respond the same. This small scale test confirms that we can assume the deformation of the individual lenses is universe away from the boundaries. }
\vspace{-0.10in}
\label{fig:abaqus}
\end{figure}

Fig. \ref{fig:abaqus} shows the results of these simulations. While there is unevenness in the stress distribution at the edge of the array, the 3 center columns and 5 center rows indicate the same deformation in response to the stretching. This small scale simulation confirms the assumption that the deformation of each lens in the array is the same so long as the individual lenses are away from the boundaries. Thus, the equations governing this system are valid and can be extended to lens arrays that have an arbitrary number of lenses due to Saint-Venant's principle. Saint-Venant's principle states that the effects of the boundary decay rapidly as the distance to the boundary increases so long as there are no sudden discontinuities in the material and the deformation is small.
 
Sudden discontinuities cause an increased amount of stress in between each lens. However, the stress concentrations are only at an extreme near the discontinuities that caused them. The impact of these discontinuities dissipates rapidly towards the center of the lens and repeats in a regular pattern. Thus, this analysis will apply to lens arrays of arbitrary resolution. Further, the boundary between each lens could be designed such that there is a smooth transition between each spherical lens portion instead of a knife edge like transition. 
 
In order to examine the optical properties of our design, we conducted a simulation of a deformable lens array viewing a tiled USAF-1951 target. We simulate a 2 mm thick lens array with 0.555 mm wide lenses. The lens array is 145 by 145 lenses and is focused at 250 mm. The target is approximately 68 mm on each side. Nine copies of the target are placed in a three by three configuration. These nine copies are then placed at the lens array's focal plane, 250 mm away from the front of the lens. We first capture the scene in the unstretched case, then we stretch the lens array by 0.2775 mm two times. This is done by finding where the primary ray for each individual lens in the array intersects with the scene. We then take the average RGB value within a region that is the size of the imaged patch as given by Eq. (\ref{eq:patch_size}). The average value in this region is then taken as the value captured by that lens' corresponding pixel. We use the same simulation to find the MTF (Modulation Transfer Function) of Stretchcam for the unstretched case, the case where the elastic lens array is stretched by 0.2775 mm and the case where the elastic lens array is stretched by 0.555 mm. 
 
\begin{figure*}[t]
\begin{center}
\includegraphics[width=0.50\linewidth]{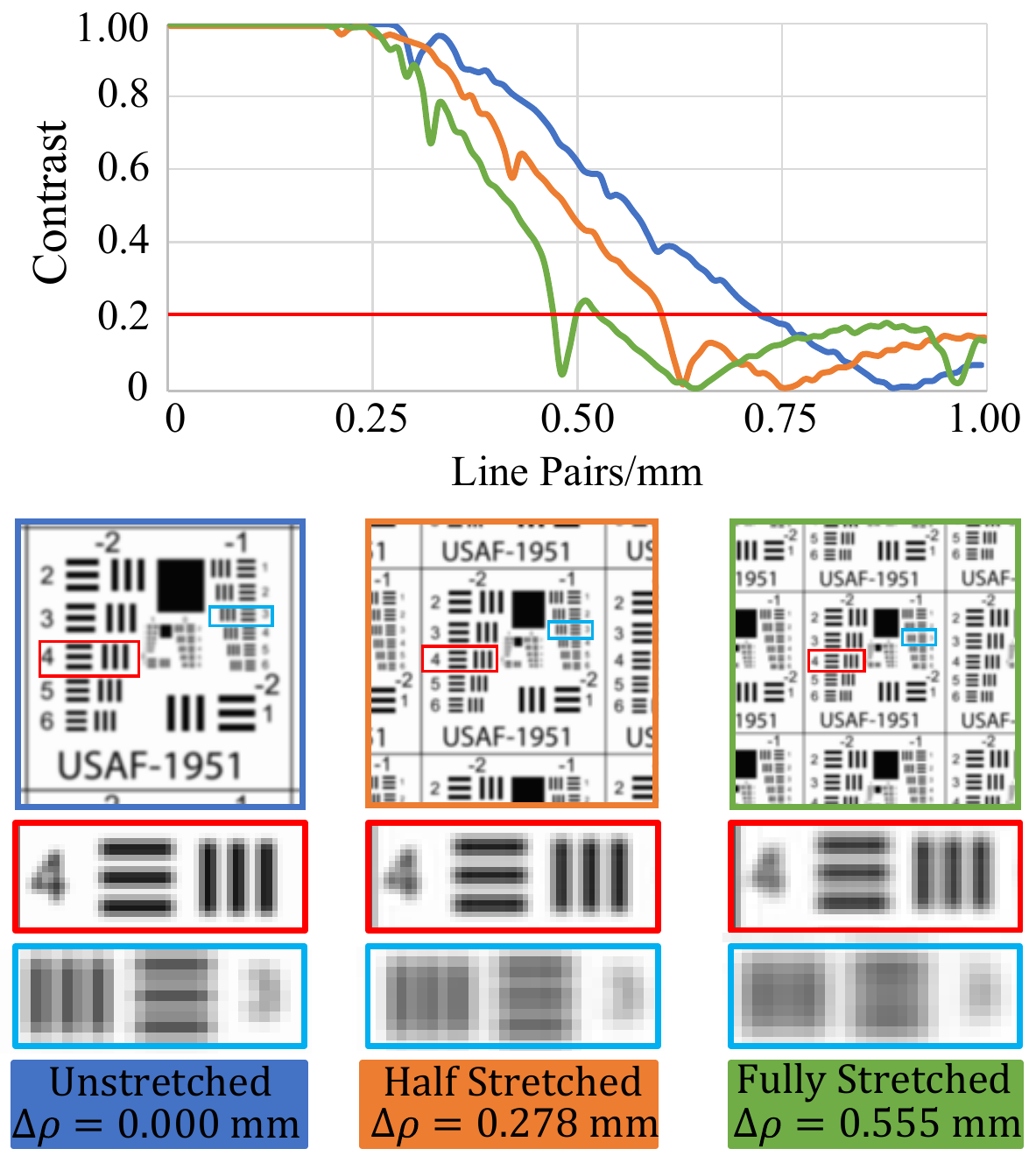}
\end{center}
\vspace{-0.15in}
\caption{This figure. shows the frequency response of the lens array at the designed focal plane, 250 mm. Notice how some line pairs remain distinguishable at all stages of the lens array's deformation while others are blurred. This is due to the decreasing Nyquist frequency of the system as the lens is stretched. USAF-1951 targets are provided that give a visual representation of the same data. Despite some frequencies falling under the contrast threshold, there is a set of frequencies common to all deformation cases that is captured with acceptable contrast. This version of Stretchcam is able to achieve a two times zoom with only a 0.555 mm stretch and is able to capture 0.354 lp/mm throughout the stretch.}
\vspace{-0.25in}
\label{fig:MTF}
\end{figure*}
 
Fig. \ref{fig:MTF} shows the results of these simulations. Notice dark blue line indicating the MTF plot for the unstretched case. While there is a drop in the contrast when the width of a line pair is equal to the width of four lenses, the Stretchcam is able to capture frequencies at 20\% contrast or greater between 0.00 lp/mm (line pairs per mm) and 0.75 lp/mm . However, as the lens array is stretched, the ability of the Stretchcam to capture higher frequencies is reduced. This is an expected behavior since, as the lens array is stretched, the shifting of the primary rays of the lenses in the array decreases the Nyquist frequency in accordance to Eq. (\ref{eq:Nyquist_Frequency}). Stretchcam also uses the same number of pixels to image larger and larger portions of the scene as the lens array is stretched, which contributes to this reduction in image quality. This reduction in the maximum frequency resolved is observed in the MTF plots for $\Delta \rho = 0.278 mm$ (indicated with an orange line) and $\Delta \rho = 0.555 mm$ (indicated with an green line) in Fig. \ref{fig:MTF}.

\begin{figure*}[t]
\begin{center}
\vspace{0.15in}
\centerline{\includegraphics[width=0.80\linewidth]{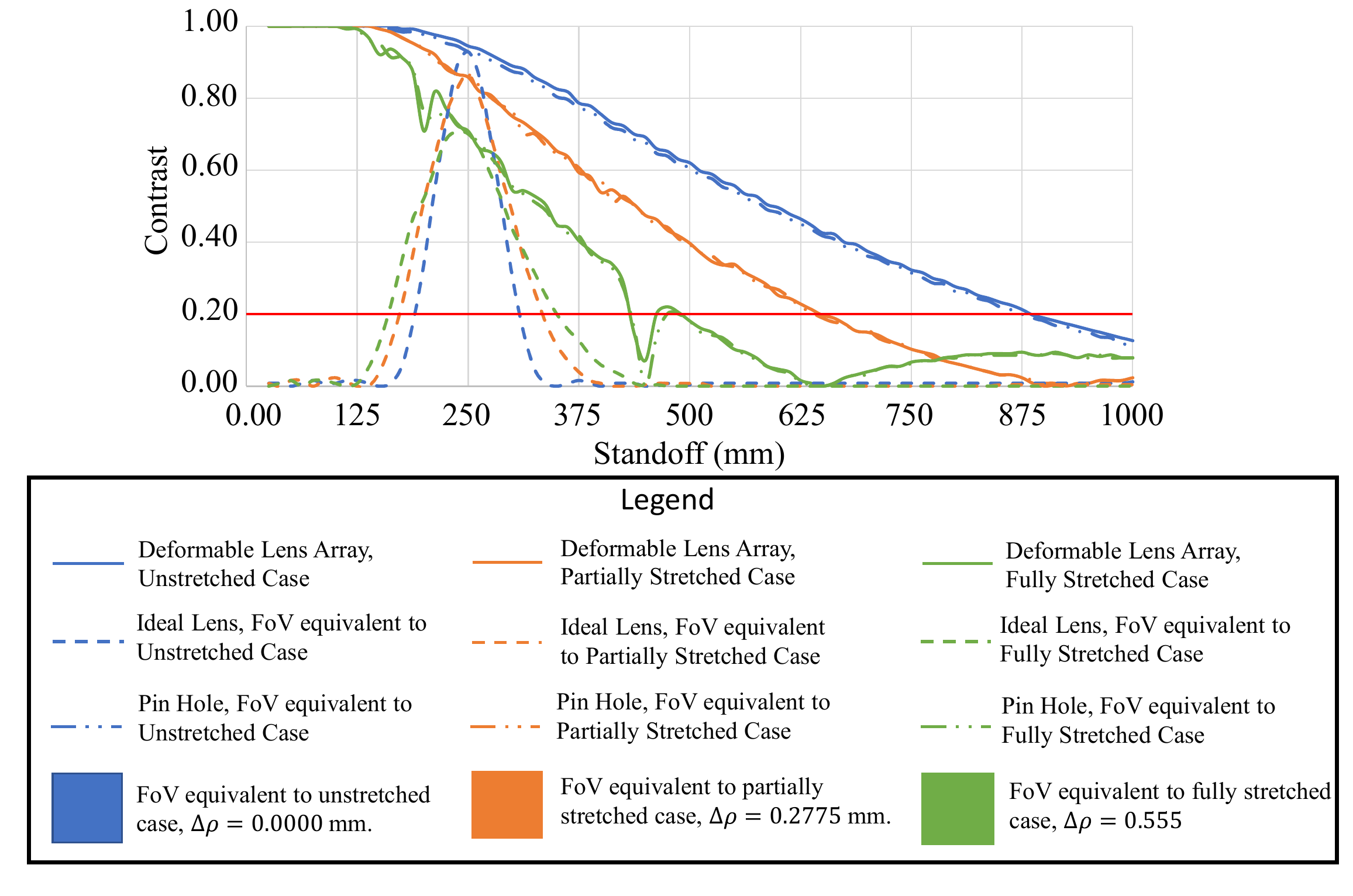}}
\hspace{-0.50in}
\end{center}
\vspace{-0.40in}
\caption{This Fig. shows the response to 0.354 lp/mm for an ideal lens, a pinhole and Stretchcam at various stand-off distances. The simulated optical systems are designed such that they capture the same field of view as Stretchcam and are all focused at 250 mm. Stretchcam behaves the same as a pin hole camera, both the pin hole camera and Stretchcam have deep near field fields of view. }
\vspace{-0.20in}
\label{fig:mtf_depth_of_field}
\end{figure*}
 
In order to study the depth of field, we conduct the same simulation for various standoff distances and plot the response for a particular frequency, in this case 0.354 lp/mm. The results of these simulations are shown in Fig. \ref{fig:mtf_depth_of_field}. Stretchcam's response is shown as solid lines for the unstretched (Dark Blue Line), partially stretched (Orange Line) and fully stretched (Green Line) cases. The contrast improves as the standoff distance decreases due to the reduced impact of diffraction blur and the increased Nyquist frequency for closer standoff distances in the stretched cases. Stretchcam has a depth of field starting at the lens surface and ending at the contrast cutoff for the desired frequency in the fully stretched case should there be no aperture. For this particular design, the depth of field ranges from 0 mm to 400 mm.
 
We compare the depth of field of our design to a conventional lens (dashed line) and a single pinhole (dashed/dotted line) in order to determine the usefulness of Stretchcam as a zoom lens. Stretchcam has a larger depth of field than a normal lens, a depth of field similar to a pinhole camera. The conventional lens and the single pinhole both have sensors arrays of 145 by 145 pixels with each pixel 1 micron by 1 micron. The sensor is placed behind the lens in such a way that the conventional lens and the pinhole will have the same field of view as the lens array. The conventional lens has its focal length set such that the lens will be in focus at a standoff distance of 250 mm given the previously mentioned distance between the lens and the sensor. The aperture diameter of the conventional lens is then set such that the lens array and the conventional lens have the same f-number. The diameter of the pinhole is such that the scene will be in best focus at 250 mm. As shown in Fig. \ref{fig:mtf_depth_of_field}, the Stretchcam has the benefits of a single pinhole camera and a much larger depth of field than a conventional lens as well as the ability to change its field of view.
 
\section{Experimental Results}
 
\hspace{0.3cm} We built a prototype to test the imaging performance of Stretchcam experimentally. Using the fabrication procedures outlined in Sims et al. \cite{Sims16}, we made a 33 by 33 lens array out of Momentive Silopren 7005, a type of extremely soft liquid silicone rubber suited to deformable optics. In order to emulate the final camera, we placed the silicone lens on a 2mm thick aluminum sheet with a rectangular grid of apertures with diameters of 1 mm. The pitch of the grid is 7 mm to match the pitch of the lens array in the undeformed case. An Optigrafix\texttrademark \hspace{0.1cm} DFMM diffusing sheet is attached to the bottom of the aperture sheet in order to capture the 33 by 33 images created by the lens array. An LCD monitor is placed above the prototype Stretchcam at the designed focal plane, 300 mm, in order control the scene displayed to the prototype Stretchcam.
 
\begin{figure}[t]
\begin{center}
\includegraphics[width=0.6\linewidth]{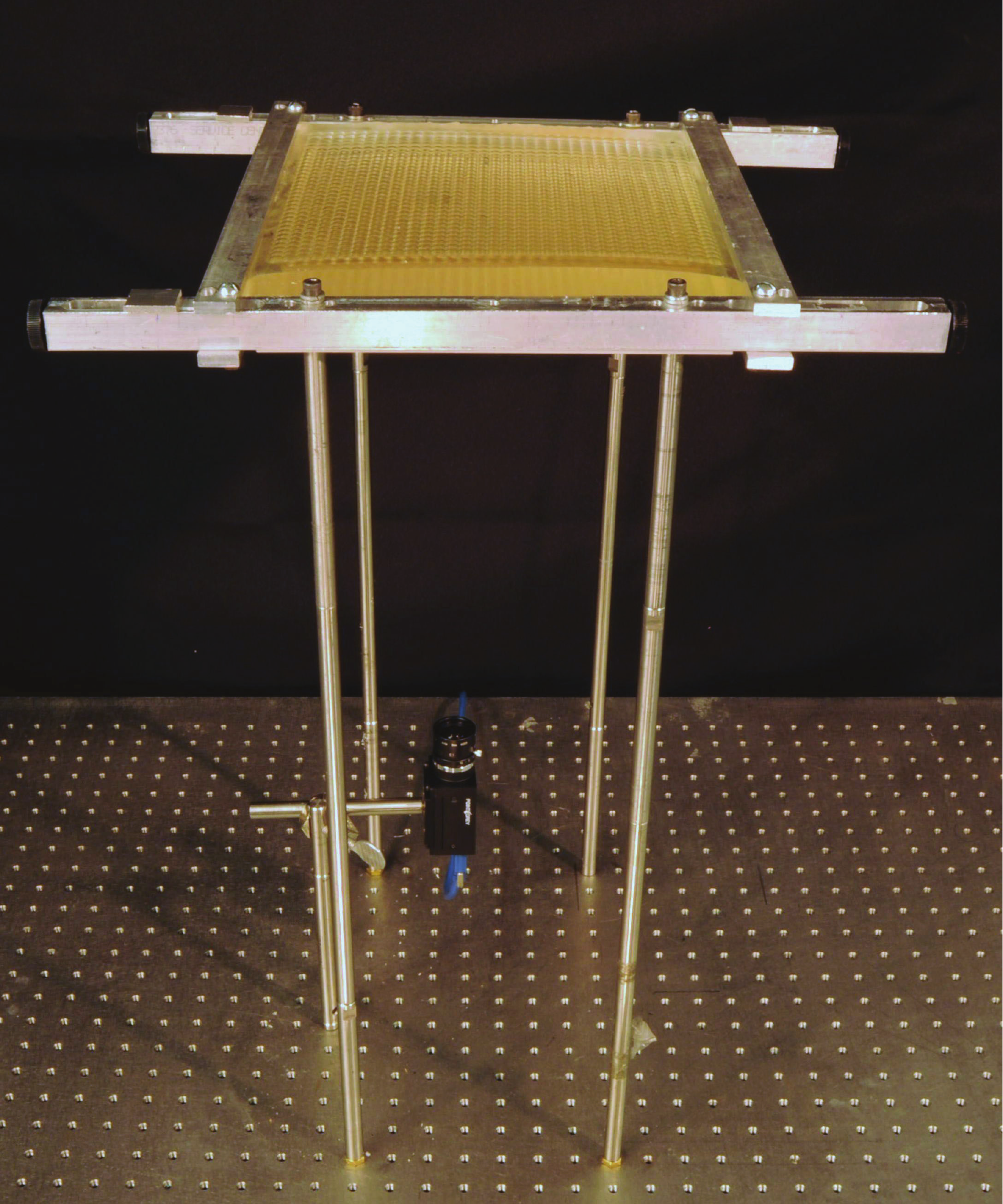}
\end{center}
\vspace{-0.15in}
\caption{Experimental apparatus. The stretching of the lens array is controlled using a spreader-like mechanism with two clamps on either side of the lens array. Images formed on the diffuser attached to the bottom of the sheet camera are captured using a digital camera and processed to produce images of scenes shown on the display. The prototype camera has a resolution of 33 by 33.}
\vspace{-0.1in}
\label{fig:rig}
\end{figure}

Fig. \ref{fig:rig} shows the prototype Stretchcam used to conduct our experiments. Two metal bars provide rails on which a spreader mechanism is placed. Attached to either end of the slider mechanism there is a clamp that is used to secure one end of the elastic lens array. Hand cranks are included to control the amount the spreaders are displaced.
  
We use a Point Grey Grasshopper 3 to capture images of the diffuser. The prototype Stretchcam only captures 33 by 33 bright spots when a white image is shown on the monitor. We save the location of these spots in the image for when various other scenes are shown to the prototype. We stretch the lens array by rotating each hand crank half a turn after which we capture more images of the diffuser. We repeat this process until the center of either row of lens furthest from the center of the lens array is aligned with the edges of the aperture plate, indicating that distance between the two extreme optical axes had increased by 7 mm.
 
%

\begin{figure*}[t]
\begin{center}
\includegraphics[width=0.65\linewidth]{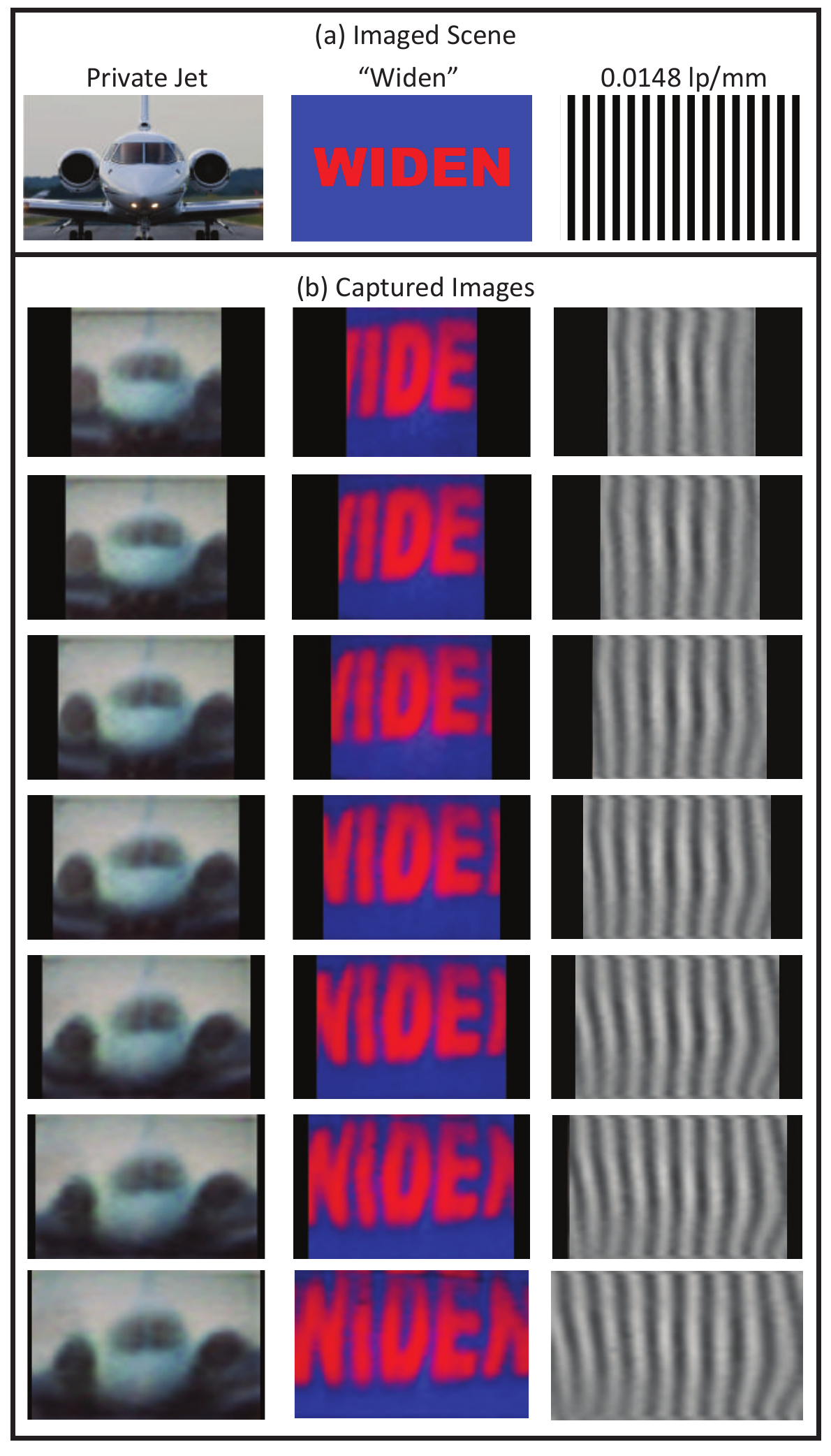}
\end{center}
\vspace{-0.15in}
\caption{ Experimental Results.  (A) Imaged scenes. These three images - Private Jet scene, ``Widen'' text scene and a Ronchi ruling with a frequency of 0.0148 lp/mm -  were displayed on a monitor above the Stretchcam prototype. (B) Captured Images. This is an sequence of the images captured by the prototype Stretchcam as lens array is stretched. The aspect ratio of these images were adjusted using interpolation to account for the increasing captured region which starts at 231 mm and ends at 333 mm. The lens array was only stretched 3\% of its original length to achieve 1.5 times zoom with each scene 300 mm away.  }
\vspace{-0.0in}
\label{fig:results}
\end{figure*}

Fig. \ref{fig:results}(a) depicts the scenes shown to the prototype Stretchcam and Fig. \ref{fig:results}(b) shows the image sequences we captured with the experimental apparatus. The lens array is deformed in such that $\Delta \rho = 7 mm$. The 33 by 33 resolution source images captured by Stretchcam are interpolated to 300 by 300 pixels. The images are then resampled to a new aspect ratio using the known field of view of the lens array. The prototype Stretchcam is able to capture a frequency of 0.0148 lp/mm with a contrast of 32\% in the unstretched case and 26\% in the stretched case. The system is able to achieve 1.5 times zoom with only a 3\%  change of the lens array's original length. The zoom value for the prototype system could be larger if designed for a scene that was further away. However, space restrictions in our lab limited the standoff distance for this test to 300 mm.



\section{Discussion}
 
\hspace{0.3cm} We have presented the design of an elastic lens array that can zoom large amounts with a small deformation. Not only can Stretchcam zoom, but it preserves the desirable depth of field aspects of a lens array. We have explored the frequency response of the system via simulation and shown there is a set of frequencies that is captured for all deformations. We have demonstrated that the depth of field of our system is not only similar to a pinhole camera but has a set of common depth of fields for all deformations that allows for a scene to remain in focus throughout the deformation. We then verified our design with an experiment and not only found that we could capture a scene without aliasing artifacts or blurring, but that we could achieve 1.5 times zoom by stretching the lens array 7 mm, which is an approximately 3\% elongation in length of the lens array. Friction did not hinder to stretching of the lens array. 

The main drawback of our design is the limited resolution and zoom capability. Lu et al. \cite{Lu2012} can achieve two times zoom with a 5.04-megapixel sensor while our system only achieves 1.5 times zoom and can capture a 0.2 Megapixel image. However, our system's ability to resolve objects is well above the threshold for identification as described by Johnson's Law\cite{Donohue}. This makes Stretchcam suitable to low power, low-cost surveillance systems. Further, the deformable mirrors in Lu et al.'s work require power to remain actuated whereas our system could take advantage of a non-back-drivable mechanism to zoom then stay in place without power. Our system is also extremely cheap, with some silicone rubbers costing only \$250 per liter. This is compared to the cost of deformable MEMS mirrors which can run in the thousands. Light \cite{Light} can cost up to \$1,699. While it is thin, its price makes it not suited to the task of a disposable and ubiquitous surveillance system.

In our experiments, we assumed that the stretching was even throughout the lens. However, one could design a system where different sections of the lens array were stretched different amounts. A user of such a device could pinch a region on their screen to cause a corresponding actuation in the elastic lens array and thus zoom in on a region of interest while retaining a wide field of view. In such a design, stress sensors will be needed in order to compute the geometry of the sheet. One could the same apply techniques that integral imagers use to enhance their viewing angle, such as elemental lens switching \cite{Lee:02}, to overcome $N_{C1}$, the resolution limit due to crosstalk. In such a case, adjacent lens could be blocked such that the lens could be stretched more. 

\section*{Acknowledgments} 

This work was supported in part by ONR grant \#N00014-16-1-2152.

This paper benefited from the input of CAVE lab members, particularly Brian Smith and Jason Holloway. Stretchcam was prototyped using the facilities of The Columbia Laboratory for Unconventional Electronics, led by John Kymissis.

{\small
\bibliographystyle{ieee}
\bibliography{egbib}
}

\end{document}